\documentclass[runningheads]{llncs}

\usepackage{graphicx}
\usepackage{hyperref}
\usepackage[all]{hypcap}

\begin{document}

\title{BrainPainter v2: Mouse Brain Visualization Software}

\author{Vedvayas Mallela\inst{1} \and Polina Golland\inst{2} \and Răzvan V. Marinescu\inst{2}}
\authorrunning{Mallela et al.}

\institute{Northview High School, Johns Creek, USA, GA 30097 \and Computer Science and Artificial Intelligence Laboratory, Massachusetts Institute of Technology, Cambridge, USA, MA 02139}

\maketitle
\begin{abstract}
BrainPainter is a software for the 3D visualization of human brain structures; it generates colored brain images using user-defined biomarker data for each brain region. However, BrainPainter is only able to generate human brain images. In this paper, we present updates to the existing BrainPainter software which enables the generation of \emph{mouse} brain images. We load meshes for each mouse brain region, based on the Allen Mouse Brain Atlas, into Blender, a powerful 3D computer graphics engine. We then use Blender to color each region and generate images of subcortical, outer-cortical, inner-cortical, top and bottom view renders. In addition to those views, we add new render angles and separate visualization settings for the left and right hemispheres. While BrainPainter traditionally ran from the browser (https://brainpainter.csail.mit.edu), we also created a graphical user interface that launches image-generation requests in a user-friendly way, by connecting to the Blender backend via a Docker API. We illustrate a use case of BrainPainter for modeling the progression of tau protein accumulation in a mouse study. Our contributions can help neuroscientists visualize brains in mouse studies and show disease progression. In addition, integration into Blender can subsequently enable the generation of complex animations using a moving camera, generation of complex mesh deformations that simulate tumors and other pathologies, as well as visualization of toxic proteins using Blender’s particle system.

\end{abstract}

\section{Introduction}
Visualizations of brain structures help researchers and medical professionals interpret data collected during experiments and trials. Highlighting differences in biomarkers across brain regions, and showing disease progression, is vital for understanding biological mechanisms. Resources for visualization of human brains are abundant with softwares such as Freesurfer (Fischl (2012)), FSLView (Jenkinson et al. (2011)), 3DSlicer (Fedorov et al. (2012)) and Statistical Parametric Mapping (SPM) (Penny Et al. (2011)) which visualize single scans, and even population averages, for a given biomarker. However, these softwares are designed for human brains and are not able to generate mice brain images. Moreover, they don’t directly offer access to an API, so it is generally cumbersome to generate a series of images repeatedly, as well as animations. In addition, they use proprietary formats, so it is often difficult to interface with them without using their respective pipeline. 

BrainPainter (Marinescu et al. (2019)), a software for template-based brain visualization, was recently proposed to address the need for repeated brain image generation, based on a generic input format. BrainPainter is capable of taking biomarker input data and generating renders of those colored regions via Blender, a 3D graphics engine. BrainPainter is an open-source visualization software that is centered around giving an intuitive experience to users who are working on illustrating brain structures.

While all the aforementioned software can visualize human brains, none of them can visualize mouse brain images. A software that could visualize mice brains would be extremely beneficial for understanding and interpreting results from studies involving mouse models, particularly on neurodegenerative diseases such as Alzheimer’s disease, Parkinson’s disease, or Frontotemporal Dementia. Due to the lack of softwares for mouse brain visualization, many studies such as Michno et al. (2019) have included the raw, tabular data in place of renders. This restricts the potential of understanding that readers could have. Although there are some extensions for these softwares such as SPM Mouse (Sawiak et al. (2013)) which enable users to register mice brains, there is a lack of up to date softwares for mouse brain visualization. Softwares such as the Allen Mouse Brain Explorer (Oh et al. (2011)) can visualize 3D mouse brains, but cannot generate renders, show animations, or color regions. 

In this work, we propose BrainPainter v2, a new extension that can render images of mouse brains. We use the Allen Mouse Brain Atlas (Lein et al. (2007)) and the Allen Mouse Brain Explorer (Oh et al. (2011)) to model mouse brains on Blender. We demonstrate this feature of our software by visualizing tau pathology progression using the data from Henderson et al. (2020). In addition, we add a Tkinter graphical user interface (GUI) that enables users to run the software straight from their computers. We further add left and right hemisphere settings to all brain atlases on BrainPainter, which lets users show asymmetries in their trials. We additionally add top and bottom views of the brain to show the disparities between regions in the left and right hemispheres. These new viewing angles generate renders that help users show how a disease is affecting a hemisphere differently from another. Through our work, we created a mouse brain visualization tool that enables neuroscientists to visualize, understand and share their work more efficiently.

\section{Methods}

\subsection{BrainPainter v1 features}
BrainPainter v1 is capable of generating human brain renders and can create animations for pathology progression (Marinescu et al. (2019)). It supported three brain atlases: Desikan-Killiany atlas (Desikan et al. (2006)), the Destrieux atlas (Destrieux et al. (2010)) and the Tourville atlas (Klein and Tourville (2012)). It could generate three types of images: an (i) outer-cortical view, an (ii) inner-cortical view, and a (iii) subcortical view, all from the right hemisphere.

BrainPainter runs from the source code (github.com/razvanmarinescu/brain-coloring), a docker container, and from a website (brainpainter.csail.mit.edu).

\begin{figure}
\centering
\includegraphics[width=\textwidth]{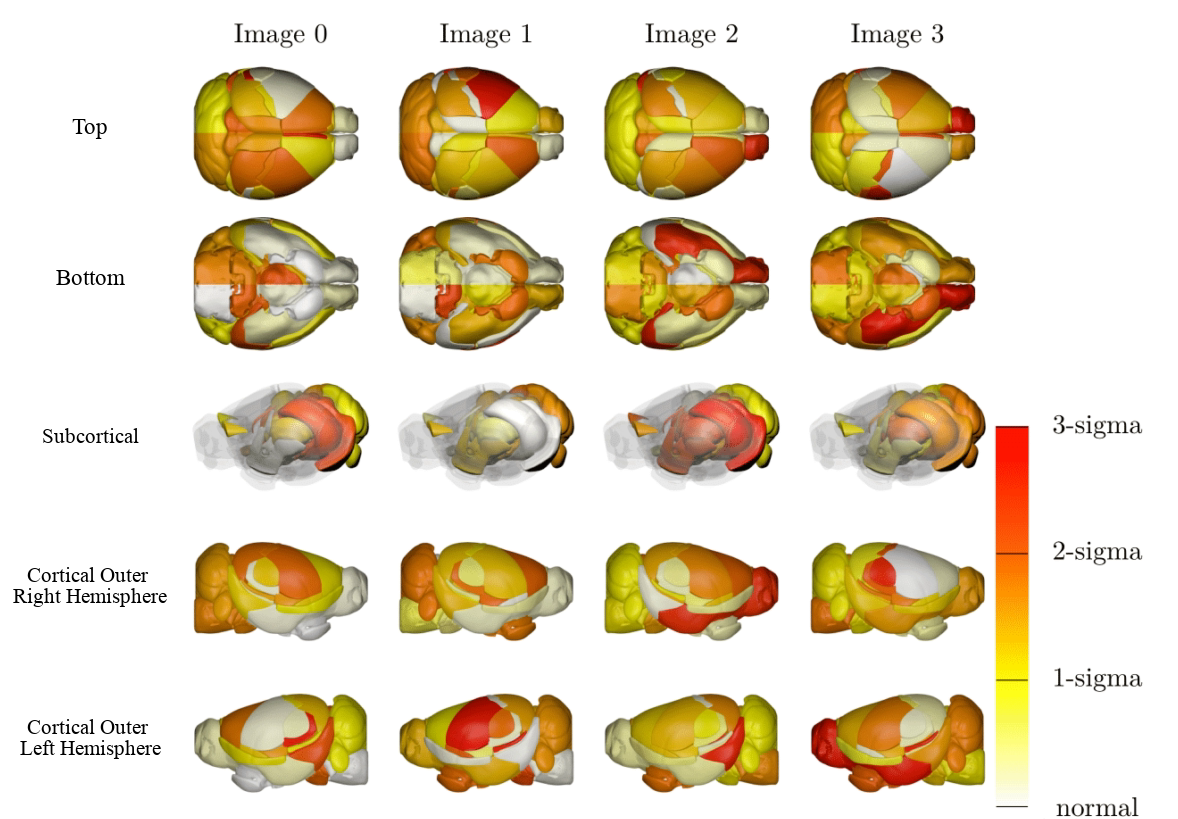}
\caption{Demonstration of BrainPainter's mouse brain visualization. This visualization uses meshes from the Allen Mouse Brain Atlas (Lein et al. (2007)) to model the mouse brain and renders the images using Blender.} \label{fig1}
\end{figure}

\subsection{Mouse Brain Visualization}
To extend BrainPainter for mouse brain visualization, we used the Allen Mouse Brain Atlas (Lein et al. (2007)) and the Allen Mouse Brain Connectivity Atlas (Oh et al. (2011)). The atlas contained 1328 regions of interest (see the full list at http://api.brain-map.org/api/v2/structure\_graph\_download/1.json) as binary masks in .ply format, which we transformed to form a mouse brain using Blender. We scaled and rotated the models from the Allen Mouse Brain Atlas (Lein et al. (2007)) so that they are on the same coordinate system as Blender. Then, we aligned the individual brain regions in relation to each other to form a full mouse brain. This model was then exported as polygon files and split into the left and right hemispheres of the atlas, enabling users to show differences between hemispheres. In Fig. \ref{fig1}, we show example images for mice brains using randomized biomarker data.

\begin{figure}
\centering
\includegraphics[width=\textwidth]{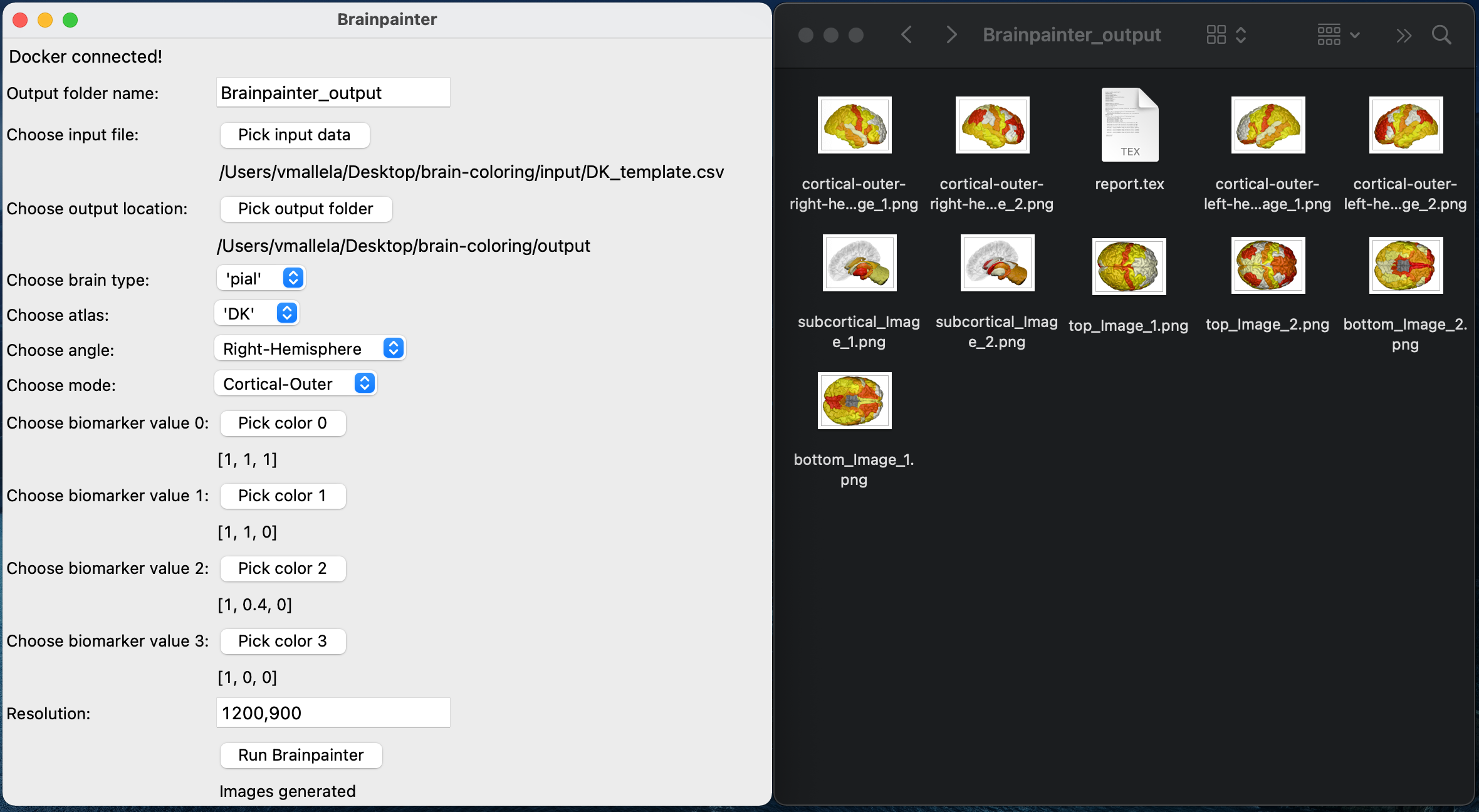}
\caption{We built a desktop GUI for BrainPainter, based on the Tkinter framework, which intuitively generates the images. The application loads the user preferences into a configuration file and passes the image generation requests to the Blender engine running in the background. To remove the need for installing Blender and its dependencies, we provide Blender pre-installed in a Docker container.} \label{fig2}
\end{figure}
\subsection{BrainPainter Graphical User Interface}
While BrainPainter can run both from the source code as well as from the browser, these often have several issues. Installing from the source code can be troublesome, because of the need to install Blender dependencies for its Python scripting language. For less technically capable users, this could lead to problems. The website is the most convenient way for users to render their images, but it is not capable of generating a large number of images in a single run. 

To overcome the aforementioned problems, we developed a graphical user interface Fig. \ref{fig2}. It (https://github.com/vmallela0/Brainpainter-GUI), uses Tkinter and the docker API to carry users' requests and retrieve images. The GUI gets information from users including the input data, brain type, atlas, render angle, resolution, and the biomarker colors. With this information, it launches a BrainPainter docker container and populates a configuration file with all of the user-defined settings. Once the configuration file is set and the docker container runs, the application retrieves the docker output and places it in an output folder on the client's machine.

\begin{figure}
\includegraphics[width=\textwidth]{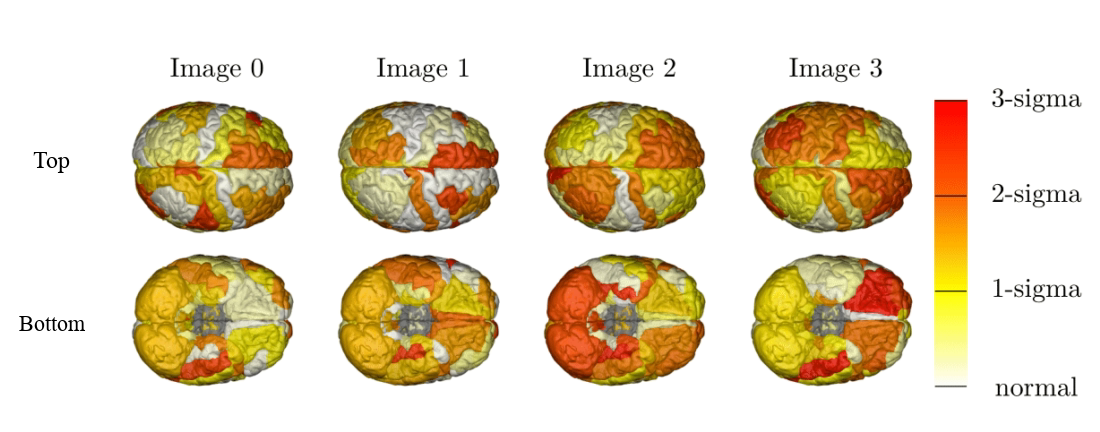}
\caption{Demonstration of BrainPainter's left and right hemisphere visualization feature. This feature enables users to show differences in how diseases might affect the two hemispheres differently.} \label{fig3}
\end{figure}

\subsection{Visualization of Left and Right Hemispheres}
To enable users to visualize asymmetries between the brain hemispheres, we extended BrainPainter to enable visualization of both hemispheres and added additional viewing angles to show asymmetries. The left hemisphere often needs to be visualized, especially for pathologies that affect one hemisphere differently than the other. Our new extension can show renders for both the left and the right sides of the brain (see Fig. \ref{fig3}), by loading both of the hemispheres' meshes into Blender. To show differences between the right and left hemispheres together, we also added a “top” and “bottom" view, which show both hemispheres in one image. This setting is available for the four atlases that BrainPainter supports: the Desikan-Killiany atlas (Desikan et al. (2006)), the Destrieux atlas (Destrieux et al. (2010)), the Tourville atlas (Klein and Tourville (2012)) and the Allen Mouse Atlas (Lein et al. (2007)). Users can define left and right hemispheres in the .csv input data that they provide. This feature is especially useful when contrasting and showing asymmetric effects of brain pathology, which have been shown for example in Alzheimer’s disease and related diseases (Lehmann et al. (2011)).

\section{Use Cases}
\begin{figure}
\centering
\includegraphics[width=\textwidth]{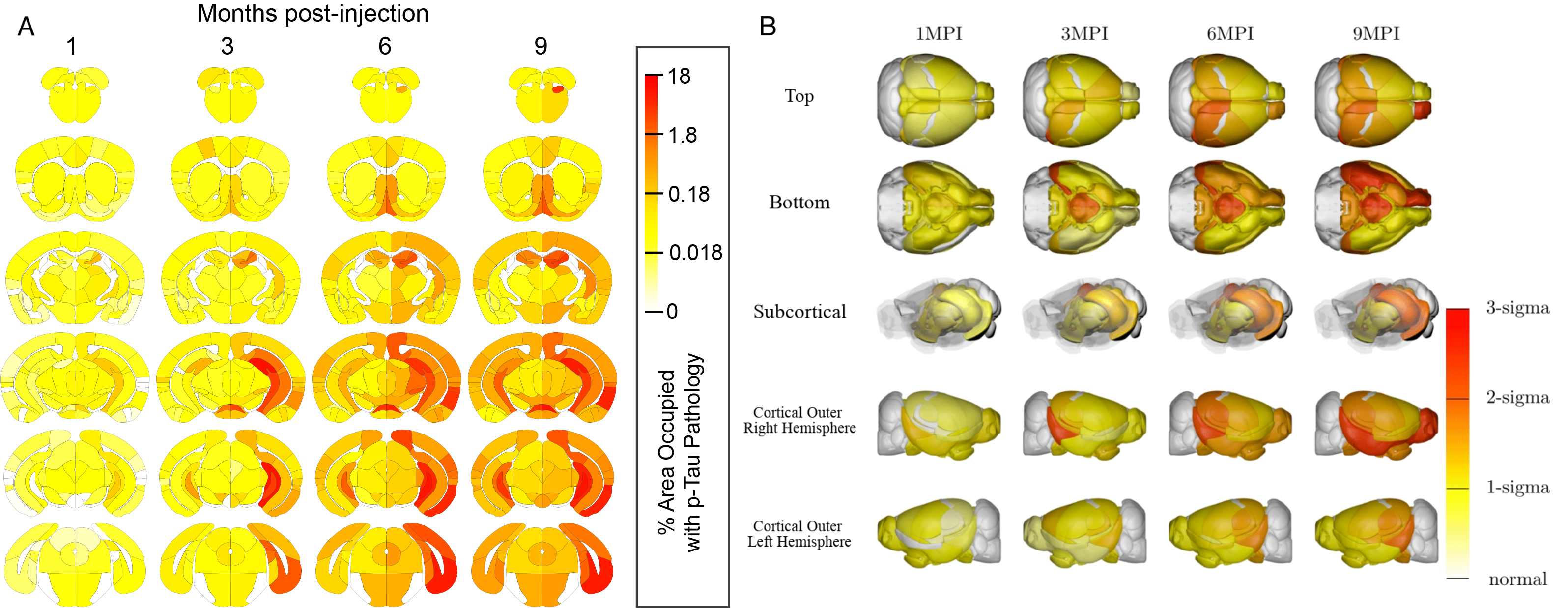}
\caption{(A) Example images from HeatMap, an in-house software at the University of Pennsylvania Center for Neurodegenerative Disease Research. (B) BrainPainter renders of the same tau pathology that HeatMap renders in A. These images show a tau pathology progression for Henderson et al. (2020} \label{fig4}
\end{figure}
\subsection{Visualising pathology progression in mice}
Visualizing the progression of pathology in neurodegenerative diseases such as Alzheimer’s is key for understanding their underlying mechanisms and finding drug targets. In Fig \ref{fig4}B, we demonstrate BrainPainter v2 on visualizing the progression of tau toxicity in mouse brains, using data from Henderson et al. (2020). Each column in Fig \ref{fig4}B shows the mouse brain in intervals of months post-injection of the tau pathogenic protein. The data used to generate these images is log-transformed over a 1000-fold difference and shows the spread of the tau pathogenic protein from Henderson et al. (2020). In the trials, BrainPainter illustrates how the right hemisphere's hippocampus and olfactory region intensifies progressively. In Fig \ref{fig4}A, we show a similar visualization using in-house software HeatMap. This differs from BrainPainter, as it produces inner slices of brains, while BrainPainter is capable of rendering images of volumetric projections. In addition, our software is open-source. Images like these can help researchers efficiently communicate their findings and can provide a visual of the complicated processes being represented by the raw data. 

\subsection{Animation of Tau Pathology Progression in Mice}
Animating the progression of pathologies, especially neurodegenerative diseases such as Alzheimer's, can help illustrate disease spread more clearly. Videos showing pathological progression can show how diseases spread through regions and can visualize diseases over time. We animate the pathology progression of the tau toxicity from Henderson et al. (2020) to show how the disease spreads over time in videos from the top and subcortical viewing angles (see supplementary). Using the ImageMagick software, we transform multiple renders into a gif of the pathology. Compared to showing multiple renders as seen in Fig 4B, videos are capable of showing how disease spreads frame by frame and leads to a better understanding of the pathology. Studies such as Thomas et al. (2014) use videos to highlight pathologies with estimated years onset to communicate their studies more effectively. Videos showing pathology progression in mice would be valuable to studies because they can further illustrate the progression seen in renders.

\section{Conclusion}
In this paper, we introduced updates to the existing BrainPainter software. By enabling visualization of mice brains, we are helping neuroscientists working with mouse brains present their findings in more understandable ways. The mouse brain visualization addition to BrainPainter will aid in generating images to interpret raw data from experiments. With the addition of visualization for the left and right hemispheres, we can now show asymmetries present in brain diseases. 

BrainPainter still has constraints that could increase usability. A problem that technically capable researchers would have while running BrainPainter from the source code is installing dependencies on Blender. BrainPainter currently runs on Blender version 2.79 which does not support pip. Since Blender has an embedded python interpreter inside the software, installing any dependencies is difficult. Updating the Blender version would cause problems with the shading techniques we use, however, because they have since been deprecated. In future versions of BrainPainter, we plan on developing a workaround that can let users experiment with the code easily in an up to date version of Blender. 

A limitation that we will address in the future is the lack of support for inner cortical views on mouse brains. The meshes we use to build mice brains are hollow and when split into left and right hemispheres, they show a hollow interior. With some transformations with Blender, we could generate meshes that not only support inner views of the brains but also can display a view from the coronal section. Another limitation of BrainPainter is its inability to color individual points/voxels on the brain mesh. However, that would require specialized input data and would increase the complexity of the software. 

\section{Acknowledgments}
VM was involved with this project through the Northview High School Directed Studies program and the Talented and Gifted program. We would like to especially thank Mr.Michael Martin from Northview High School for his support. 

RVM was supported by the NIH grants NIBIB NAC P41EB015902, NIH NINDS U19NS115388, as well as IBM and Takeda.

We are also thankful to Dr.Michael Henderson from the Van Andel Institute for his help with obtaining data for the tau pathology and for providing examples of other existing mouse brain visualization tools.

\end{document}